\documentclass{article}
\usepackage{spconf,amsmath,graphicx}

\usepackage{booktabs}
\usepackage{multirow}
\usepackage{hyperref}
\usepackage[table,xcdraw]{xcolor}
\usepackage{svg}

\newcommand{\DIS}{\texttt{DIS}}
\newcommand{\WER}{\texttt{WER}}

\newcommand{\MAE}{\texttt{MAE}}
\newcommand{\AAE}{\texttt{AAE}}
\newcommand{\old}{\texttt{old}}
\newcommand{\new}{\texttt{new}}
\newcommand{\Southern}{\texttt{Southern}}
\newcommand{\TopSeventy}{\texttt{Top70}}

\title{Improving Speech Recognition for African American English With Audio Classification}
\name{\parbox{.95\linewidth}{\centering
Shefali Garg, Zhouyuan Huo, Khe Chai Sim, Suzan Schwartz, Mason Chua, Al\" ena Aks\" enova, \\
Tsendsuren Munkhdalai, Levi King, Darryl Wright, Zion Mengesha, Dongseong Hwang, \\ Tara Sainath, Fran\c coise Beaufays, Pedro Moreno Mengibar}}
\address{Google LLC}
\begin{document}
\ninept
\maketitle
\begin{abstract}
Automatic speech recognition (ASR) systems have been shown to have large quality disparities between the language varieties they are intended or expected to recognize.  One way to mitigate this is to train or fine-tune models with more representative datasets.  But this approach can be hindered by limited in-domain data for training and evaluation.
We propose a new way to improve the robustness of a US English short-form speech recognizer using a small amount of out-of-domain (long-form) African American English (AAE) data.  We use CORAAL, YouTube and Mozilla Common Voice to train an audio classifier to approximately output whether an utterance is AAE or some other variety including Mainstream American English (MAE).  By combining the classifier output with coarse geographic information, we can select a subset of utterances from a large corpus of untranscribed short-form queries for semi-supervised learning at scale.  Fine-tuning on this data results in a $38.5\%$ relative word error rate disparity reduction between AAE and MAE without reducing MAE quality.
\end{abstract}
\section{Introduction}
The goal of ASR systems is to transcribe speech, allowing voices from speakers of many accents and language varieties to be understood. Diverse speech recognition has increasingly become the focus of researchers and policy makers. For example, France recently debated outlawing discrimination by accent\footnote{\url{https://www.bbc.com/news/world-europe-55069048}}; and some researchers investigate how the bias affecting certain groups of speakers arises in speech recognition models, and how to mitigate this bias~\cite{markl2022language}.
Literature reporting on the results of various model evaluations shows that ASR models often perform better for monolingual MAE speech than, for example, AAE~\cite{doi:10.1073/pnas.1915768117}, utterances with code-switching~\cite{vielzeuf2022e2e}, or speech in areas with less MAE prevalence~\cite{9053751}. In turn, speakers who are consistently mis-recognized might try decreasing their speech rate or accommodating their linguistic features towards the recognizer's biases in order to be understood~\cite{DHERAM_2022,mengesha2021impact}. This is why measuring bias and finding ways to improve ASR models is critical.
Dheram et al. \cite{DHERAM_2022} present an initial study to identify commonly misrecognized cohorts and mitigate the performance bias. They show that geographic location can act as an indicator to identify misrecognized cohorts but is not precise enough. In our paper, we show how we can combine coarse geographic information along with an dialect classifier model to select data for the target missrecognized cohort. We focus on African American English (AAE) as our target cohort. AAE is used daily by over 30 million speakers~\cite{ewave} across the United States. AAE has phonological systems, vocabulary, grammatical constructions, and other linguistic factors that differ from Mainstream American English (MAE) and might not be well-recognized if models are MAE-centric. Koenecke et al. \cite{doi:10.1073/pnas.1915768117} showed that the Google's ASR quality for Black users was around $1.6$ times worse than for white users.

\label{sec:intro}
Our contribution is to use dialect classification for data selection to measure and improve the quality disparity of an American English speech recognizer for AAE compared to MAE.
We train the classifier (section \ref{sec:dialect_classification}) to detect AAE audio using a small amount of long-form (LF) AAE speech from the Corpus of Regional African American Language (CORAAL) (section \ref{sec:coraal}) and YouTube as positive training examples and utterances from YouTube and Common Voice as negative training examples. The classifier generalizes well to our short-form (SF) domain, with a precision of $89$\% and a recall of $94.7$\% when evaluated on anonymized Voice Search data with human-provided dialect labels.
We then use the classifier and coarse location data to mine AAE-specific and non-AAE utterances and transcribe them with a teacher model~\cite{narayanan2019recognizing} for ASR training (section~\ref{sec:asr-data-selection}). This corpus allows us to fine-tune an end-to-end speech recognizer model, reducing its relative AAE-specific WER by 7.7\% and its relative disparity against AAE by 38.9\%.  We also use a matched n-gram study (section \ref{sec: matched_n_gram}) similar to \cite{doi:10.1073/pnas.1915768117}, which shows a relative WER disparity reduction of 48.7\% on common English words spoken in AAE-classified utterances relative to non-AAE-classified utterances. This demonstrates an improvement in acoustic modeling and not just text prediction.
    
To the best of our knowledge, this framework of using audio classification for AAE to reduce WER disparity is novel and aims towards fairer production-scale ASR systems.
\begin{figure}[t]
  \centering
  \includegraphics[width=\linewidth]{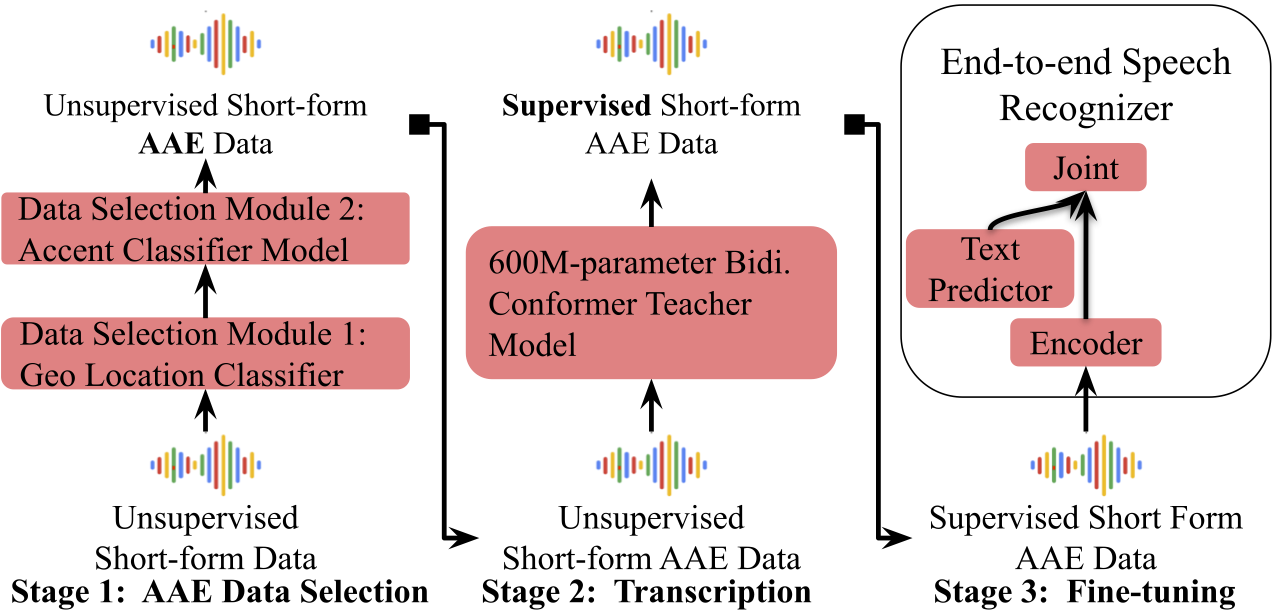}
  \caption{Framework to reduce the WER disparity between AAE and MAE for any ASR system where we have limited and out-of-domain supervised data for AAE.} 
  \label{fig:framework}
\end{figure}

\section{AAE Classification}

The goal of the dialect classifier is to output $1$ for AAE speech and $0$ for non-AAE varieties of English, including MAE.  It is composed of a pre-trained speech foundation model~\cite{li2023efficient} and a 2-layer fully-connected network classifier with Swish activation~\cite{ramachandran2017searching}. The foundation model is a $2$-layer convolutional network followed by a $24$-layer conformer~\cite{gulati2020conformer} encoder with hidden dimension $1024$ and about $600$M parameters in total. The input is a log-Mel feature vector of size $128$. The foundation model was pre-trained on a multilingual YouTube dataset using the BEST-RQ~\cite{chiu2022self} algorithm for $800$k steps.

We use roughly $20$ hours of CORAAL~\cite{kendall2018corpus} \label{sec:coraal} which is a corpus of long-form sociolinguistic interviews with AAE speakers in three US communities. For the binary classification task, we use Common Voice US data~\cite{commonvoice:2020} as negative examples. Additionally, $18.03$ hours of MAE YouTube and $22.61$ hours of AAE Youtube data are also used to train the classifier. To increase training speed and alleviate the mismatch with our target short-form domain, we segment the long recordings randomly into utterances with length from 5 to 20 seconds. 

There are three non-trivial factors considered for the dialect classifier: (1) pooling method, (2) classifier depth and (3) training some foundation model layers. For simplicity, we evaluate precision and recall of compared methods on the dev set at threshold $0.5$.

In order to combine the encoder outputs into a fixed-size input for the fully-connected network, a pooling method is inserted between them. We compare three methods: average pooling, maximum pooling, and attentional pooling~\cite{lee2019set} where the query vector is from average pooling by training on Common Voice US and CORAAL data. The classifier is a 2-layer fully connected network and all parameters are updated in this subsection. Results in Table~\ref{tab:pool} show that the best quality can be achieved by using the average pooling.

\begin{table}[t]
  \caption{Pooling method for the encoder outputs. Evaluated on Common Voice US and CORAAL.} 
  \label{tab:pool}
  \centering
\setlength{\tabcolsep}{2.5 pt}
\begin{tabular}{@{}ccc@{}}
\toprule
{ Pooling} & { Precision} & {Recall} \\
\midrule 
Average  & \textbf{ 94\%} & { 94\%} \\
Maximum  & { 93\%} & { 93\%}  \\
Attentional  & { 91\%}  & \textbf{ 95\%} \\
\bottomrule
\end{tabular} 
\end{table}

We manually optimized the depth of the fully-connected network between 1 and 3, finding that two layers are better than one or three in Table~\ref{tab:fcn}. In the experiment, average pooling is used and all parameters are updated. 

\begin{table}[t]
  \caption{Number of layers for the fully-connected network. Evaluated on Common Voice US and CORAAL.}
  \label{tab:fcn}
  \centering
\setlength{\tabcolsep}{2.5 pt}
\begin{tabular}{@{}ccc@{}}
\toprule
{ $\#$ layers} & { Precision} & {Recall} \\
\midrule 
1  & { 82\%} & {93\%} \\
2  & \textbf{ 94\%} & \textbf{ 94\%} \\
3  & { 91\%} & { 94\%}  \\
\bottomrule
\end{tabular} 
\end{table}

Fine-tuning all parameters of the pre-trained speech foundation model could lead to over-fitting, given that the amount of classifier training data is small compared to the amount of parameters. In this subsection, we experiment with updating only a small part of the parameters. In table~\ref{tab:partial}, we fine-tune the bottom, middle or top two layers in addition to training the 2-layer fully-connected network on Common Voice US, CORAAL, and YouTube MAE and AAE audio. It is obvious that fine-tuning the middle two layers show better performance than compared methods.  
\begin{table}[t]
  \caption{Updated partial foundation model. Evaluated on MAE and AAE YouTube.}
  \label{tab:partial}
  \centering
\setlength{\tabcolsep}{2.5 pt}
\begin{tabular}{@{}ccc@{}}
\toprule
{ Trainable layers} & { Precision } &{Recall} \\
\midrule 
All  & { 68\%} & {100 \%}\\
Layers 0,1  &{ 65\%} & {95 \%} \\
Layers 11,12  & \textbf{ 88\%} & \textbf{100 \%} \\
Layers 22,23  & { 76\%} & {100 \%} \\
\bottomrule
\end{tabular} 
\end{table}

\begin{table}[t]
  \caption{Dialect classifier results on target SF data.}
  \label{tab:dialect_classifier}
  \centering
\setlength{\tabcolsep}{2.5 pt}
\begin{tabular}{@{}ccc@{}}
\toprule
{ Test set} & { Precision} & { Recall} \\
\midrule
{ SF Full set } & { 79.4\%} & { 30.6\%} \\
{ SF Verified Dev set (200 utts)} & { 84.7\%} & { 37.1\%} \\
{ SF Verified Test set (100 utts)} & { 89.0\%} & { 94.7\%} \\
\bottomrule
\end{tabular} 
\end{table}
\label{sec:dialect_classification}

\section{AAE Data Selection}
\label{sec:asr-data-selection}
We use short-form AAE data as the target domain and language variety for our experiments. We select this data on the basis of geographic information and the dialect classifier.

\subsection{Geographic Locations}
\label{geographic-locations}
As a comparison and supplement to the dialect classifier, we use some coarse geographic features as a loose proxy for language variety.  We create the \TopSeventy\ set containing utterances from the $70$ most populated metropolitan areas in the United States, and the \Southern~set containing utterances from southern regions of United States with a high prevalence of AAE~\cite{Aksenova2022AccentedSR}.  The \Southern\ corpus is likely to contain more data that is sociophonetically similar to AAE data, including consonental and vocalic characteristics common to AAE, while the \TopSeventy\ set is likely to include more MAE data due to overlap with MAE-prevalent regions like the Midwest, Central Plains, Rocky Mountains and West Coast~\cite{wolfram2015american}.

\subsection{Dialect Classifier Scores}
\label{sec:classifier-thresholds}
Since language variety cannot be determined by location alone \cite{DHERAM_2022}, we use the dialect classifier to further filter the \TopSeventy\ and \Southern\ sets. 
We manually optimize the threshold for classifying the utterances using the SF Verified Dev set, shown in Table \ref{tab:dialect_classifier}. Classifying the utterances to be AAE when the classifier score is $\ge 0.7$ and MAE when $< 0.4$ gave us a precision and recall of 89\% and 94.7\% on SF Verified Test Set. We use this threshold to further select the AAE data.

\begin{figure}[t]
  \centering
  \includegraphics[width=.8\linewidth]{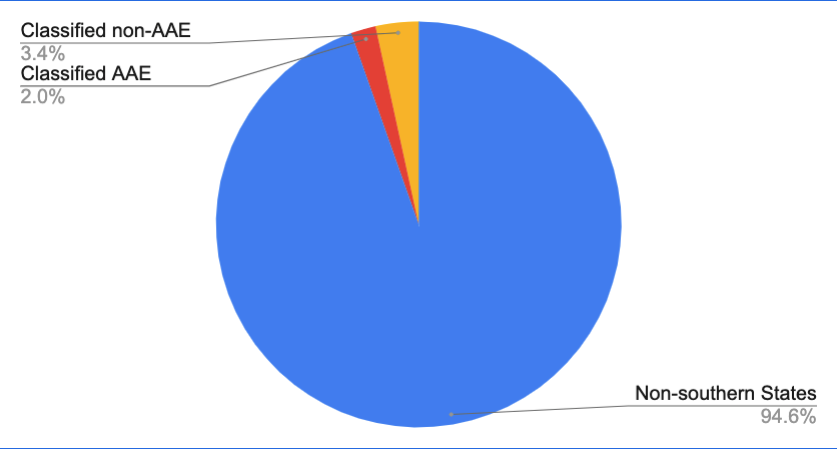}
  \caption{Distribution of original short-form training data. Shows the small proportion of AAE data in our existing training set.}
  \label{fig:limited_train}
\end{figure}
Fig.~\ref{fig:limited_train} shows the original distribution of short-form supervised data without mining any new AAE data. Only $5$\% of our original dataset contains data from the southern states. Out of this, only $2$\% of the data was categorized as AAE by the classifier. This clearly shows a lack of representation of data containing AAE-like features in our current training set. 

\begin{table}[t]
  \caption{AAE training data statistics selected using dialect classifier and location proxy.}
  \label{tab:data_stats}
  \centering
\setlength{\tabcolsep}{1.3 pt}
\begin{tabular}{@{}lcc@{}}
\toprule
{ \begin{tabular}[c]{@{}l@{}}Classifier-selected \\ AAE Training
Data\end{tabular}} & \# Utterances & \begin{tabular}[c]{@{}c@{}}\% of Total \\ SF Dataset\end{tabular} \\
\midrule
{Unsupervised SF \Southern} & { 2.8M} & { 0.6\%} \\
{Unsupervsied SF \TopSeventy} & { 5.6M} & { 1.2\%} \\
\bottomrule
\end{tabular}
\end{table}
Table~\ref{tab:data_stats} shows that using the region, and dialect classifier scores, we mined AAE data equivalent to around $1.8$\% of the total original unsupervised short-form training data.

\label{sec:data_selection}
\section{Experimental Setup}

\subsection{Model Architecture Details}
The ASR model used in these experiments is a 120M-parameter hybrid autoregressive transducer (HAT) \cite{variani2020hat}. Its encoder has 12 conformer~\cite{gulati2020conformer} layers with model dimension 512. As the model is streaming ASR, we restrict it from using any future information \cite{li2021better}. The convolution kernel size is 15, and the self-attention layer has 8 heads with a left-context length of 65.
The model input is a vector of size 528, consisting of 4 contiguous frames of 128-dimensional log-Mel features \cite{narayanan2019recognizing} sub-sampled by a factor of 3. The text labels are coded with a word piece model~\cite{37842} with 4,096 vocabulary items.
\label{sec: model_architecture_details}

\subsection{Data Sets}
\label{sec: data_sets}
The base model, before any fine-tuning with AAE, was trained with an English voice-search short-form dataset composed of anonymized utterances transcribed by people (supervised) or by a $600M$-parameter bidirectional conformer teacher model (semi-supervised). See~\cite{narayanan2019recognizing} for details on the base model and its teacher. \\
For the AAE datasets, we use the \Southern\ AAE and \TopSeventy\ AAE created using the coarse geographic locations and classifier scores as described in section \ref{sec:data_selection}. We also partition them into train and test sets.

Our primary AAE train and test set is \Southern\ AAE, since by construction we expect \Southern\ to contain a larger proportion of AAE utterances than \TopSeventy.  For the primary MAE test set, we select data from MAE-rich United States regions with a classifier score $< 0.4$ as described in section \ref{sec:data_selection}. We use this set as a proxy for MAE quality, and call it ``MAE'' for brevity, even though the classifier was trained to assign low scores to all non-AAE varieties, not just MAE. We use the same process to create our secondary test sets, \TopSeventy\ AAE and MAE. All utterances are anonymized. The test utterances were human-labeled, and the training utterances were teacher-labeled.

\subsection{Evaluation Metrics}
\label{sec: evaluation_metrics}

\subsubsection{AAE/MAE WER Disparity}
Disparity (\DIS) is the relative increase in word error rate (WER) on the AAE test set compared to the MAE test set.
\begin{equation} \label{eq1}
\begin{split}
\textbf{\DIS} & = \frac{\WER(\AAE) - \WER(\MAE)}{\WER(\MAE)} \\
\end{split}
\end{equation}
We measure the relative disparity reduction between 2 ASR models as follows:
\begin{equation} \label{eq2}
\begin{split}
\textbf{Disparity Reduction} & = \frac{\DIS(\old) - \DIS(\new)}{\DIS(\old)} \\
\end{split}
\end{equation}
\label{disparity}
\subsubsection{Matched N-Gram WER}
\label{sec: matched_n_gram}
Ideally, the performance of ASR on AAE speech should be evaluated on two datasets based on the same transcriptions to eliminate lexical differences between the datasets and focus only on acoustic differences. However, our existing test sets do not meet such requirements. 
To circumvent this problem, we explore the “matched n-gram” evaluation, similar to the one used in~\cite{doi:10.1073/pnas.1915768117}

Given the reference transcripts from two datasets, $T_1$ and $T_2$, we can find the matched n-grams pairs as follows: 
\begin{enumerate}
\item Extract the n-grams $N_1$ and $N_2$ from the transcripts, $T_1$ and $T_2$, limiting the order as desired (e.g. to $2$ and $3$ in our case).
\item Find the common n-grams $N$ $=$ $N_1 \cap N_2$.
\item For each n-gram $N$, find its corresponding utterances from the two datasets, $\{ u_1, \ldots, u_L \}$ and $\{ v_1, \ldots, v_M \}$.
\item Create paired utterances $\{ (u_1, v_1), (u_2, v_2),$ $\ldots, (u_P, v_P) \}$ where $P = \min(L, M)$.
\item Compute the standard Levenshtein edit alignment between each hypothesis and its reference transcript for WER computation.
\item Discard the alignment edges that are not part of the utterance's matched n-gram in the reference, then perform the standard WER computation on the remaining alignments (i.e. correct words divided by total words).
\end{enumerate}
For evaluating our AAE model on matched N-grams, we extract common bi-grams and tri-grams from our primary test set. We extracted a total of 2567 common n-grams out of which 1074 were unique. We report WER on this acoustic-focused test set for our experiments.

\section{Experimental Results}
\label{sec: experimental_results}
In this section, we conduct extensive dialect adaptation experiments to improve the model's WER on AAE test sets while maintaining its performance on MAE test sets.

\subsection{AAE/MAE WER Disparity Reduction}
We trained our model on the SF dataset and fine-tuned it on AAE datasets. Table~\ref{tab:finetune_aae_primary} shows the performance of our baseline recognizer on our primary test set. In Row $1$, when no AAE data was used for fine-tuning, we saw a relative WER disparity of $25$\%. 
We fine-tuned the recognizer using the new semi-supervised AAE SF data (row 2), which reduced the AAE WER from 6.5 to 6.2. However, MAE WER regressed slightly. To overcome overfitting, we reduced the learning rate of the model, which not only resolved the MAE set regression but also further improved the AAE WER. Using this AAE data, for our best model, we were able to reduce the MAE/AAE disparity to $15.4$\% from $25$\%, achieving a relative disparity reduction of $38.9$\%.
\begin{table}[t]
  \caption{ASR fine-tuned on AAE evaluated with the primary test set (\Southern\ and \MAE).}
  \label{tab:finetune_aae_primary}
  \centering
\setlength{\tabcolsep}{2.5pt}
\begin{tabular}{ccccccc}
\toprule
\multicolumn{2}{c}{Data} & Learning & \multicolumn{2}{c}{WER} & MAE/AAE & Disparity \\
Train & Fine-tune & Rate & MAE & AAE & Disparity & Reduction \\
\midrule
{ SF} & \begin{tabular}[c]{@{}c@{}}None\end{tabular} & 7.5 & 5.2 & 6.5 & 25.0\% & - \\
SF & SF AAE & 7.5 & 5.3 & 6.2 & 17.0\% & 32.0\% \\
{SF} & {SF AAE} & { 5.0} & 5.3 & 6.2 & 17.0\% & 32.0\% \\
SF & SF AAE &\textbf{ 2.5} &\textbf{ 5.2} & \textbf{6.0 }& \textbf{15.4\%} &\textbf{ 39.0\%} \\
\bottomrule
\end{tabular}
\end{table}
Table~\ref{tab:finetune_aae_secondary} shows the same metrics for our secondary test set, \TopSeventy, including a relative disparity reduction of $15.1$\%.

\begin{table}[t]
  \caption{ASR fine-tuned on AAE evaluated with the secondary test set (\TopSeventy~\AAE/\MAE). }
  \label{tab:finetune_aae_secondary}
  \centering
\setlength{\tabcolsep}{2.5pt}
\begin{tabular}{ccccccc}
\toprule
\multicolumn{2}{c}{Data} & Learning & \multicolumn{2}{c}{WER} & MAE/AAE & Disparity \\
Train & Fine-tune & Rate & MAE & AAE & Disparity & Reduction \\
\midrule
{ SF} & { \begin{tabular}[c]{@{}c@{}}None\end{tabular}} & { 7.5} & { 4.3} & { 6.6} & { 53.5\%} & { -} \\
{ SF} & { SF AAE} & { \textbf{2.5}} & { \textbf{4.4}} & { \textbf{6.4}} & { \textbf{45.4\%}} & { \textbf{15.1\%}} \\
\bottomrule
\end{tabular}
\end{table}

\subsection{Acoustic-specific Improvements}
To estimate the improvement due to gains on dialects alone, we used the matched n-gram test to compare the WER of the baseline model on AAE and non-AAE audio of the same content. Table $8$ shows that the baseline model has a low WER, as expected due to its construction with common bi- and tri-grams across language varieties. However, there is still a large quality disparity, with a WER of 1.2\% for non-AAE audio and 2.1\% for AAE audio. Fine-tuning on the new AAE corpus reduces this disparity by 48.7\%.

\begin{table}[t]
  \caption{Matched n-gram evaluation.}
  \label{tab:n-gram}
  \centering
\setlength{\tabcolsep}{2.5pt}
\begin{tabular}{@{}ccccc@{}}
\toprule
\multirow{2}*{Model} & \multicolumn{2}{c}{WER} & MAE/AAE & Disparity \\
& MAE & AAE & Disparity & Reduction \\
\midrule
{ Baseline} & { 1.2} & { 2.1} & { 75.0\%} & { -} \\
{ Our Model} & { 1.3} & { \textbf{1.8}} & { \textbf{38.5\%}} & { \textbf{48.7\%}} \\
\bottomrule
\end{tabular}
\end{table}

\subsection{Impact of the Geographic Proxy}
To study the benefit of coarse region information, we fine-tuned the base model with AAE training data from the \TopSeventy\ and \Southern\ training sets separately. Table $7$ shows that fine-tuning with the \Southern\ AAE data alone reduced the disparity to $10.9$\%, while fine-tuning with the \TopSeventy\ AAE data gave us a disparity of $16.7$\%. This suggests that location proxy plays an important role in selecting high-quality AAE data. We also observed that the \TopSeventy\ dataset has almost double the number of utterances ($5.5$ million vs. $3$ million) as \Southern, yet we saw a higher improvement with \Southern. This shows that simply adding more data does not improve AAE WER. Instead, we need to carefully select data based on location and dialect classifier scores to reduce the disparity between AAE and MAE WER.
\begin{table}[t]
  \caption{Fine-tuning on the \TopSeventy\ and \Southern\ training sets.}
  \label{tab:top70_vs_southern}
  \centering
\setlength{\tabcolsep}{2.5pt}
\begin{tabular}{@{}cccccc@{}}
 \toprule
Train & Tuning & $\sim$ \#Utts & \multicolumn{2}{c}{WER} & MAE/AAE \\
Data & Data & (million) & MAE & AAE & Disparity \\
 \midrule
\multicolumn{6}{l}{\textbf{\Southern\ Test Set}} \\
 \midrule
{ SF} & { None} & { 0m} & { 5.2} & { 6.5} & { 25.0\%} \\
 
{ SF} & { All SF AAE} & { 8.5m} & { 5.2} & { 6.0} & { 15.5\%} \\
 
{ SF} & { \TopSeventy\ SF AAE} & { 5.5m} & { 5.4} & { 6.3} & { 16.7\%} \\
SF & \Southern\ SF AAE & \textbf{3m} & \textbf{5.5} & \textbf{6.1} & \textbf{10.9\%} \\
\midrule 
\multicolumn{6}{l}{\textbf{\TopSeventy\ Test Set}} \\
\midrule 
SF & None & 0m & 4.3 & 6.6 & 53.9\% \\
 
SF & All SF AAE & 8.5m & 4.4 & 6.4 & 45.5\% \\
 
SF & \TopSeventy\ SF AAE & 5.5m & 4.4 & \textbf{6.4} & 45.5\% \\
SF & \Southern\ SF AAE & 3m & 4.5 & \textbf{6.4} & 42.2\% \\
\bottomrule
\end{tabular}
\end{table}
\subsection{Wins and Losses}
\label{sec: winsandlosses}
Among utterances from our primary test set where the baseline and fine-tuned model transcripts disagree, we observed numerous ``wins'' involving rare words, as well as cases where the improved model better recognizes consonant clusters, diphthongs and word-final consonants that are pronounced according to AAE phonology; some examples are shown in Table~\ref{tab:model_wins}.
\begin{table}[t!]
  \caption{Model Wins}
  \label{tab:model_wins}
  \centering
\setlength{\tabcolsep}{2.5pt}
\begin{tabular}{cccc}
\toprule
Error & Ground & \multicolumn{2}{c}{Model Prediction} \\
Type & Truth & Baseline & Ours \\
\midrule
{\multirow{2}*{dialect} } & { foreman} & {\color[HTML]{770000} farmer} & {\color{blue} foreman} \\
{\multirow{2}*{related} } & { car} & {\color[HTML]{770000} cough} & {\color{blue} car} \\
 & clothes & {\color[HTML]{770000} close} & {\color{blue} clothes} \\
\midrule
{\multirow{2}*{Rare} } & niacin & {\color[HTML]{770000} Madison} & {\color{blue} niacin} \\
{\multirow{2}*{words} } & Ronettes & {\color[HTML]{770000} romance} & {\color{blue} Ronettes} \\
 & thy & {\color[HTML]{770000} the} & {\color{blue} thy} \\
\bottomrule
\end{tabular}
\end{table}

\section{Conclusions}
\label{sec:conclusions}
We improved a speech recognizer by reducing its WER disparity between Mainstream American English and African American English by about $38.5\%$.  To overcome the domain mismatch between the available AAE datasets and the ASR training data, we trained an audio classifier based on a foundation model and used it to create new in-domain AAE datasets for semi-supervised ASR training.  
While it remains a challenge to improve fairness of ASR systems without representative data, 
this is one step towards counteracting biases in English recognition quality.
Further directions of research include evaluating on test sets that were not selected by the classifier in order to study how well this approach generalizes to the diverse range of AAE varieties.

\bibliographystyle{IEEEbib}
\bibliography{references}

\end{document}